\documentclass[doublecol,figures]{epl2} 
\usepackage{amssymb,amsmath}
\usepackage{url}

\DeclareMathOperator\atan{atan}
\DeclareMathOperator\arctanh{arctanh}
\DeclareMathOperator\Ci{Ci}
\DeclareMathOperator\Si{Si}
\newcommand{\bea}{\begin{eqnarray}}
\newcommand{\eea}{\end{eqnarray}}
\newcommand{\be}{\begin{equation}}
\newcommand{\ee}{\end{equation}}
\newcommand{\kk}{\mathbf{k}}

\newcommand{\qq}{\mathbf{q}}

\newcommand{\OO}{\mathbf{0}}
\newcommand{\rr}{\mathbf{r}}
\newcommand{\xx}{\mathbf{x}}
\newcommand{\FF}{\hat{\psi}}
\newcommand{\ve}{\mathbf{e}}
\newcommand{\kf}{k_{\rm F}}

\title{Polaron residue and spatial structure in a Fermi gas}
\shorttitle{Polaron residue and spatial structure in a Fermi gas}
\author{Christian Trefzger and Yvan Castin}  
\shortauthor{Christian Trefzger and Yvan Castin}  
\institute{Laboratoire Kastler Brossel, \'Ecole Normale Sup\'erieure and CNRS, UPMC, 24 rue Lhomond, 75231 Paris, France}


\abstract{We study the problem of a mobile impurity of mass $M$ interacting 
{\sl via} a $s$-wave broad or narrow Feshbach resonance with a Fermi sea of particles of mass $m$.
Truncating the Hilbert space to at most one pair of particle-hole
excitations of the Fermi sea, we determine ground state properties of the polaronic branch
other than its energy, namely the polaron quasiparticle residue $Z$, and
the impurity-to-fermion pair correlation function $G(\xx)$.
We show that $G(\xx)$ deviates from unity at large distances
as $-(A_4+B_4 \cos 2 \kf x)/(\kf x)^4$, where $\kf$ is the Fermi momentum; since $A_4>0$ and $B_4>0$,
the polaron has a diverging rms radius and exhibits Friedel-like oscillations.
In the weakly attractive limit,
we obtain analytical results, that in particular detect the failure of the Hilbert space
truncation for a diverging mass impurity, as expected from Anderson orthogonality catastrophe;  
at distances between $\sim 1/\kf$ and the asymptotic distance where the $1/x^4$ law applies, 
they reveal that $G(\xx)$ exhibits an intriguing multiscale structure.}

\begin{document}
\maketitle
\date{\today}

\section{Introduction}
The physics of atomic Fermi gases has recently experienced a fast development,
thanks to the Feshbach resonance technique that allows to tune the $s$-wave scattering length $a$ of the interaction,
and to obtain highly degenerate strongly interacting Fermi gases \cite{prems,BookZwerger}.
The first realizations of spin polarized configurations~\cite{Zwierlein,Hulet} asked for theoretical interpretation 
of the experimental results, and shortly afterwards it was proposed that, in the strongly polarized case,
the minority atoms dressed by the Fermi sea of the majority atoms form a normal gas of 
quasiparticles called polarons~\cite{Lobo,Chevy},
which agrees with the experimental phase diagram~\cite{Zwierlein,Navon}.
Strictly speaking, such polarons should be called ``Fermi polarons", so as to distinguish
from the traditional condensed-matter-physics polaron, {\sl i.e.}\ an electron coupled to a bosonic bath of phonons
\cite{RevPol}.

The basic single-polaron properties, such as its binding energy to the Fermi sea and its effective mass,
are now well understood, both for broad 
\cite{Lobo,Chevy, Combescot,Svistunov} and narrow \cite{Massignan,Zhai,TrefzgerCastin} Feshbach resonances.
Here we study the polaron ground state properties going beyond the energy search, for both
types of resonances.
We study the quasiparticle residue $Z$, already investigated experimentally in \cite{Zwierlein2,Zaccanti} 
and theoretically in \cite{Zwerger,Massignan}, 
and the intriguing issue of the density distribution that surrounds the impurity,
as characterized by the pair correlation function:
This gives access to the spatial extension of the polaron, of potential important consequences 
on the properties of the polaronic gas.

\section{Model}
\label{sec:themodel}

We consider in three dimensions an ideal gas of $N$ same-spin-state fermions of mass $m$, enclosed in a cubic quantization volume $V$ with periodic boundary conditions. The gas is perturbed by an impurity,
that is a distinguishable particle, of mass $M$.
The impurity interacts with each fermion
resonantly on a $s$-wave broad or narrow Feshbach resonance,
as described by a two-channel model
\cite{Holland,Gurarie,CRAS}, with a Hamiltonian $\hat{H}$ written in \cite{TrefzgerCastin}:
The particles exist either in
the form of fermions or impurity in the open channel,
or in the form of a tightly bound fermion-to-impurity molecule in the closed channel.
These two forms are coherently interconverted by the interchannel coupling of amplitude $\Lambda$
\footnote{We can neglect the open-channel fermionic interaction
if its (background) scattering length $a_{\rm bg}$ is $\approx$ the van der Waals length.}.
We restrict to the zero-range (or infinite momentum cut-off) limit,
so that the interaction is characterized by the $s$-wave scattering length $a$ 
and the non-negative Feshbach length $R_*$ \cite{Petrov}. In terms of 
the effective coupling constant $g$ and of the interchannel coupling $\Lambda$, one has
\be
\label{eq:defct}
g = \frac{2\pi\hbar^2 a}{\mu} \ \ \ \ \mbox{and}\ \ \ \  R_*=\frac{\pi\hbar^4}{\Lambda^2\mu^2},
\ee
where $\mu=mM/(m+M)$ is the reduced mass.  A broad Feshbach resonance corresponds to $R_*=0$.

\section{Polaronic ansatz}
Whereas the ground state of the system presents two branches, a {\it polaronic} branch and a {\it dimeronic} 
branch~\cite{Svistunov,Zwerger,Mora,Combescot_bs,Massignan,Zhai,TrefzgerCastin},
we restrict to the polaronic branch.
The ground-state of the $N$ fermions is the usual Fermi sea (``FS"), of energy $E_{\rm FS}(N)$.
We determine the ground state of a single impurity interacting with the $N$ fermions using the 
unexpectedly high-accuracy approximation
proposed in \cite{Chevy} (and generalized to encompass the narrow resonance case in \cite{Zhai,TrefzgerCastin}), 
that truncates the Hilbert space to at most one pair of particle-hole 
excitations of the Fermi sea\footnote{Such a truncation in principle requires some control of its accuracy. In the weakly interacting regime, we shall validate it by comparison to a perturbative expansion. 
In the strongly interacting regime, one can add a second pair of particle-hole excitation in the ansatz, as done
in \cite{Combescot}, but this is out of the scope of the present work. One can also compare to diagrammatic 
Monte Carlo techniques \cite{Svistunov},
which show (up to now for $R_*=0$ and $M=m$) that the values of $Z$ obtained from (\ref{eq:ansatz_polaron}) 
are in excelleent agreement with the exact $Z$ even at unitarity \cite{VanHoucke}.}.
For a zero total momentum, this corresponds to the ansatz
\be
\label{eq:ansatz_polaron}
|\psi_{\rm pol}\rangle =  \bigg(\phi\, \hat{d}_\OO^\dag
+\sum'_{\qq}  \phi_\qq \hat{b}_{\qq}^\dag \hat{u}_\qq
+ \sum'_{\kk,\qq}  \phi_{\kk\qq} \hat{d}^\dag_{\qq-\kk}
\hat{u}^\dag_{\kk} \hat{u}_\qq
\bigg)
|\mbox{FS}\rangle,
\ee
where $\hat{d}_\kk^\dagger$, $\hat{u}_\kk^\dagger$ and $\hat{b}_\kk^\dagger$ are the creation operators
of an impurity, a fermion and a closed-channel molecule of wave vector $\kk$.
The prime above the summation symbol means that the sum is restricted to $\qq$ belonging to the Fermi sea
of $N$ fermions, and to $\kk$ not belonging to that Fermi sea.
The successive terms in (\ref{eq:ansatz_polaron}) correspond in that order to the ones generated
by repeated action of the Hamiltonian $\hat{H}$ on the $\Lambda=0$ polaronic ground state.
One then has to minimize the expectation value of $\hat{H}$ within the ansatz~(\ref{eq:ansatz_polaron}), 
with respect to the variational parameters $\phi$, $\phi_\qq$ and $\phi_{\kk \qq}$, with the constraint
$\langle\psi_{\rm pol}|\psi_{\rm pol}\rangle = 1$.
Expressing $\phi_{\kk\qq}$ in terms of $\phi_{\qq}$
and $\phi_{\qq}$ in terms of $\phi$, as in \cite{TrefzgerCastin},
one is left with a scalar implicit equation for the polaron energy counted with respect
to $E_{\rm FS}(N)$; in the thermodynamic limit:
\be
\label{eq:Epol}
\Delta E_{\rm pol} \equiv E_{\rm pol} - E_{\rm FS}(N) =  \int'\!\!\frac{d^3q}{(2\pi)^3} \frac{1}{\mathcal{D}_\qq},
\ee
where the prime on the integral over $\qq$ means that it is restricted to the Fermi sea $q<k_{\rm F}$,
with the Fermi momentum $\kf$ related as usual to the mean density of the Fermi sea
$\rho=N/V$ by $k_{\rm F}=(6\pi^2\rho)^{1/3}$.
The function of the energy in the denominator of the integrand is
\begin{multline}
\label{eq:defD}
\mathcal{D}_\qq=\frac{1}{g}-\frac{\mu k_{\rm F}}{\pi^2\hbar^2}
+ \frac{\mu^2 R_*}{\pi\hbar^4} \left(\Delta E_{\rm pol}+\frac{\mu}{m}\, \varepsilon_\qq\right)
\\ + \int '\!\!\frac{d^3k'}{(2\pi)^3} \left(\frac{1}{E_{\qq-\kk'}+\varepsilon_{\kk'}
-\varepsilon_\qq-\Delta E_{\rm pol}}
-\frac{2\mu}{\hbar^2 k^{'2}}\right),
\end{multline}
where $\varepsilon_\kk = \frac{\hbar^2\kk^2}{2m}$ for the fermions,
$E_\kk = \frac{\hbar^2\kk^2}{2M}$ for the impurity,
and the prime on the integral over $\kk'$ means that it is restricted to $k'>k_{\rm F}$.

\begin{figure}[t]
\begin{center}
\includegraphics[width=\columnwidth,clip=]{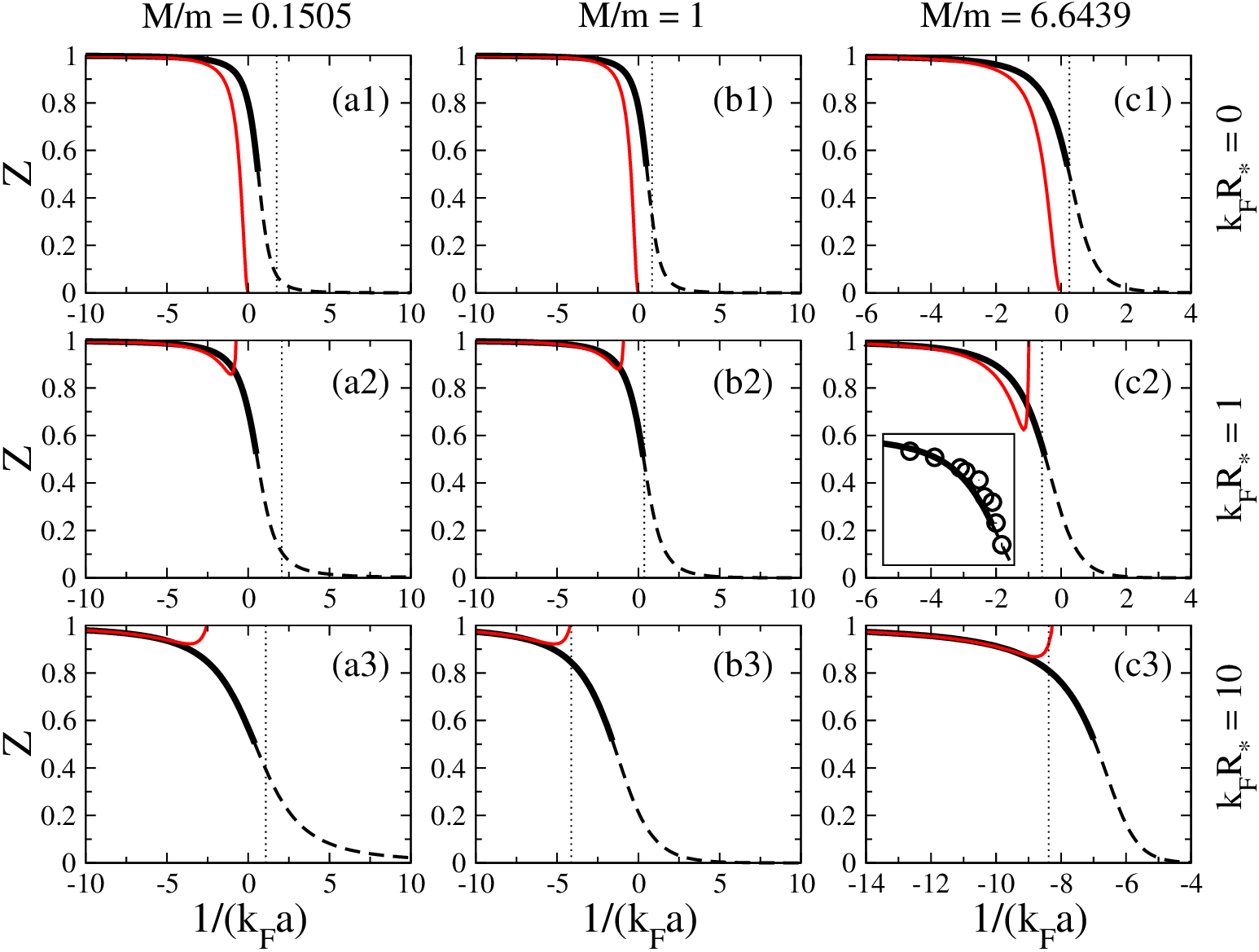}
\vspace{-0.5cm}
\caption{(Color online) Quasiparticle residue (\ref{eq:defZ}) for various mass ratios 
$M/m$ (columns a,b,c) and various Feshbach lengths (rows 1,2,3).
Black line (solid for $Z>1/2$, dashed for $Z<1/2$): Numerical solution. 
Red line: Second order weakly attractive expansion (\ref{eq:Zwal}), that diverges for $R_*>0$ 
when $s$ of Eq.~(\ref{eq:defs}) tends to unity, and for $R_*=0$ when $1/(\kf a)\to 0$.
Circles in the inset [$(\kf a)^{-1}\in [-3,0], Z \in [0.3,1]$] of (c2): 
Experimental data of \cite{Zaccanti} for $\kf R_*\simeq 0.9474$. 
Vertical dotted lines \cite{TrefzgerCastin}: 
Polaron-to-dimeron crossing point; on the left (resp.\ right) of this line, 
the ground state is polaronic (resp.\ dimeronic) [for (a1), there is
a very thin zone to the right of this line where the ground state is trimeronic \cite{Mathy}].
}
\label{fig:residue}
\end{center}
\end{figure}

\section{Quasiparticle residue}
The polaron is a well-defined quasiparticle if it has a non-zero quasiparticle residue $Z$,
which is defined in the Green's function formalism from the long imaginary-time decay
of the Green's function~\cite{Svistunov}.
Within the polaronic ansatz (\ref{eq:ansatz_polaron}), it was shown in \cite{Zwerger} that simply
$Z=|\phi|^2$.
We first write the amplitude $\phi_\qq$ in terms of the denominator (\ref{eq:defD}) 
as~\cite{TrefzgerCastin}:
\be
\label{eq:def_phi_q}
\phi_\qq = \frac{\mathcal{A}}{\mathcal{D}_\qq} 
\ee
where $\mathcal{A}$ is a normalization factor.
Then using Eq.~(\ref{eq:def_phi_q}) and the coupled equations for $\phi$, $\phi_\qq$ and $\phi_{\kk\qq}$~\cite{TrefzgerCastin}, we 
get the following expression for the residue:
\begin{multline}
\label{eq:defZ}
Z\equiv |\phi|^2 = \bigg[1+\frac{1}{\Lambda^2}\int'\frac{d^3q}{(2\pi)^3}\frac{1}{\mathcal{D}_\qq^2} \\
+ \int'\frac{d^3kd^3q}{(2\pi)^6}\left(\frac{1/\mathcal{D}_\qq}{E_{\qq-\kk}+\varepsilon_{\kk}-\varepsilon_\qq-\Delta E_{\rm pol}}\right)^2\bigg]^{-1}. 
\end{multline}
In the limit $R_*=0$, this exactly reproduces the result of \cite{Bruun} from the diagrammatic formalism
in the ladder approximation. For $R_*>0$, this can be shown to reproduce also the results of \cite{Massignan}
obtained in the ladder approximation generalized to a two-channel model.
In Fig.~\ref{fig:residue} we plot $Z$ as a function of $1/\kf a$ for various
mass ratios $M/m$ and reduced Feshbach lengths $\kf R_*$. We find that 
$Z$ tends to $1$ when $a\to 0^-$, as expected, and to $0$  when $a\to 0^+$. 
The polaronic ansatz {\sl a priori} makes sense when $Z$ is close to unity, and its accuracy
becomes questionable when $Z\to 0$. In Fig.~\ref{fig:residue}, we thus have
plotted $Z$ in dashed line for a predicted value below $1/2$. In the weakly attractive limit, we 
shall give below a systematic expansion of $Z$ up to second order in $\kf a$.

\section{Pair correlation function}

\begin{figure}[t]
\begin{center}
\includegraphics[width=0.97\columnwidth,clip=]{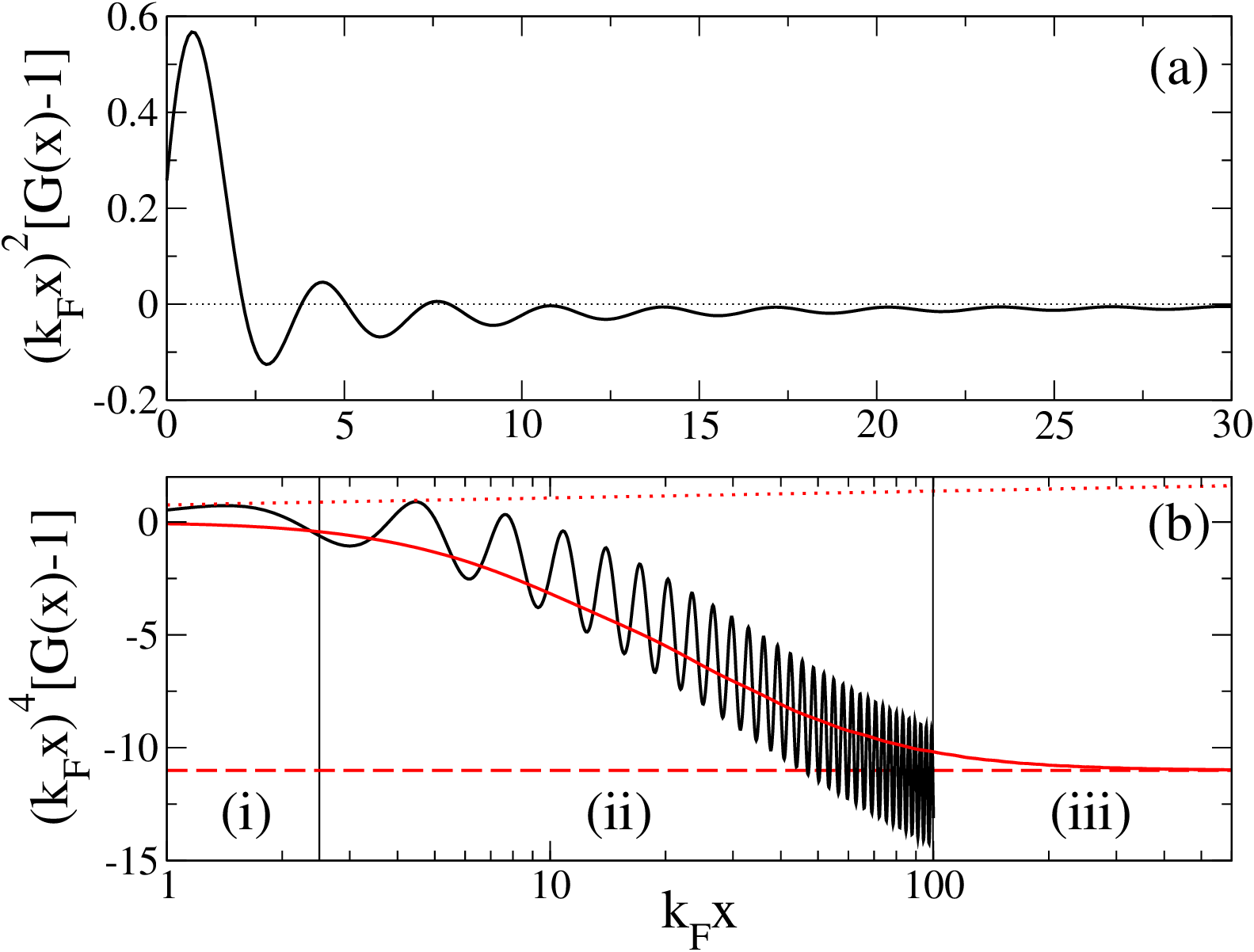}
\vspace{-0.5cm}
\end{center}
\caption{(Color online) Deviation of the pair correlation function from unity (its uncorrelated value),
for $M/m=6.6439$, $\kf R_*=1$ and $1/\kf a = -2$. (a,b) Black thick line:  Numerical solution~(\ref{eq:g2_bs}).
(b) Red solid line: Multiscale prediction~(\ref{eq:multiscale}). Dashed line: Non-oscillating
bit of the asymptotic prediction~(\ref{eq:g2large}) with $A_4$ given by~(\ref{eq:A_4_a0}).  
Dotted line: Non-oscillating bit of the prediction~(\ref{eq:limsim}). (i), (ii) and (iii):
Zones of the multiscale structure of $G(\xx)$ defined below Eq.~(\ref{eq:multiscale}).
}
\label{fig:g2}
\end{figure}

The pair correlation function $G(\xx_u-\xx_d)$ is proportional to the probability density 
of finding a fermion at position $\xx_u$ knowing that the impurity is localized at $\xx_d$.
It is thus an observable quantity, that can be extracted from a measurement of the positions of all
particles in a given realisation of the gas, which in turn has to be averaged over many 
realisations\footnote{Single shot spatial cold-atom distributions {\sl integrated} over some direction $z$ 
can now be accurately measured by absorption imaging, giving access to spatial noise
and its correlations \cite{Blochnoise}. 
The unwanted integration over $z$ can be undone by an inverse Abel transform \cite{Zwierlein3}.}.
In terms of the fermionic and impurity field
operators $\FF_u(\xx_u)$ and $\FF_d(\xx_d)$:
\be
\label{eq:defg2}
G(\xx_u-\xx_d) = \frac{\langle \FF_u^\dag(\xx_u)\FF_d^\dag(\xx_d) \FF_d(\xx_d) \FF_u(\xx_u)\rangle}{\rho \rho_d}.
\ee
Here $\rho=N/V$ is the unperturbed mean fermionic density, and $\rho_d =
\langle \FF_d^\dagger(\xx_d) \FF_d(\xx_d)\rangle$ is the mean density of impurity for the interacting system.
Due to the interchannel coupling, the impurity has a non-zero probability $\pi_{\rm closed}$
(studied in \cite{TrefzgerCastin}) to be tightly bound within a closed-channel molecule,
where it cannot contribute to $G(\xx)$ and to $\rho_d$. In terms of the probability 
$\pi_{\rm open}=1-\pi_{\rm closed}$ for the impurity to be in the open channel, 
one finds that $\rho_d =\pi_{\rm open}/V$.
In the thermodynamic limit, the open-channel probability is related to the quasiparticle residue as
\begin{multline}
\frac{\pi_{\rm open}}{Z}=
1 + \int' \frac{d^3k d^3q}{(2\pi)^6}
\left(\frac{1/\mathcal{D}_\qq}{E_{\qq-\kk}+\varepsilon_\kk-\varepsilon_\qq -\Delta E_{\rm pol}}\right)^2 
\label{eq:def_rho_d}
\end{multline}
while the pair correlation function is 
\begin{multline}
\label{eq:g2_bs}
G(\xx) = 1 + \frac{Z}{\rho\,\pi_{\rm open}}\Big[ 
-2 f(\xx) +\int'\!\!\! \frac{d^3q}{(2\pi)^3}|f_\qq(\xx)|^2 \\
-\int'\!\!\! \frac{d^3k}{(2\pi)^3}|\tilde{f}_\kk(\xx)|^2 \Big],
\end{multline}
where we have introduced the functions
\begin{eqnarray}
\label{eq:fq_def}
f_\qq(\xx) &=& \int'\! \frac{d^3k}{(2\pi)^3} \frac{e^{i\kk\cdot\xx}/\mathcal{D}_\qq}
{E_{\qq-\kk}+\varepsilon_\kk-\varepsilon_\qq -\Delta E_{\rm pol}} \\
\label{eq:fk_def}
\tilde{f}_\kk(\xx) &=& \int'\! \frac{d^3q}{(2\pi)^3} \frac{e^{i\qq\cdot\xx}/\mathcal{D}_\qq}
{E_{\qq-\kk}+\varepsilon_\kk-\varepsilon_\qq -\Delta E_{\rm pol}}\\
\label{eq:f_def}
f(\xx) &=& \int'\!\frac{d^3k d^3q}{(2\pi)^6} \frac{e^{i(\kk-\qq)\cdot\xx}/\mathcal{D}_\qq}
{E_{\qq-\kk}+\varepsilon_\kk-\varepsilon_\qq -\Delta E_{\rm pol}}.
\end{eqnarray}
Interestingly, the contribution involving $f(\xx)$ is an interference effect
between the subspaces with zero and one pair of particle-hole excitations in the Fermi sea.
In Fig.~\ref{fig:g2} we plot the numerically obtained deviation $G(\xx)-1$ of the pair correlation function from unity 
(which is both its large distance limit and its non-interacting limit)\footnote{The 
function $x^2G(\xx)$ has a finite limit
in $\xx=\mathbf{0}$ since the impurity-to-fermion wavefunction diverges as the inverse relative distance.
For $R_*=0$ and $m=M$, this limit is called the contact \cite{Tan}.}.
The presence of the impurity induces oscillations in the fermionic
density that are still significant at distances of several $1/\kf$. 

\section{Properties of the pair correlation function}
A first property of the $G(\xx)$ function is the sum rule:
\be
\int d^3x [G(\xx) - 1] = 0,
\label{eq:regsom}
\ee
where the thermodynamic limit was taken and the integral is over the whole space.
This sum rule follows directly from the integral representation of the Dirac delta distribution 
$\int d^3x\exp(i\kk\cdot\xx) = (2\pi)^3\delta(\kk)$. 

A second property is that, in the limit where $x\to+\infty$,
\be
\label{eq:g2large}
G(\xx)-1 \underset{x\to +\infty}{\sim} -\frac{A_4+B_4 \cos(2\kf x)}{(\kf x)^4}.
\ee
The prefactor of $1/(\kf x)^4$ is thus a periodic function of $x$ of period $\pi/\kf$, with a mean value $A_4$ 
and a cosine contribution (of amplitude $B_4$) reminiscent of the Friedel oscillations.  
The fact that the mean value $A_4$ differs from zero has an important physical consequence: It shows that
the polaron is a spatial extended object, since even the first moment $\langle x\rangle$ of $G(\xx)-1$
diverges (logarithmically) in the thermodynamic limit.

\begin{figure}[t]
\begin{center}
\includegraphics[width=\columnwidth,clip=]{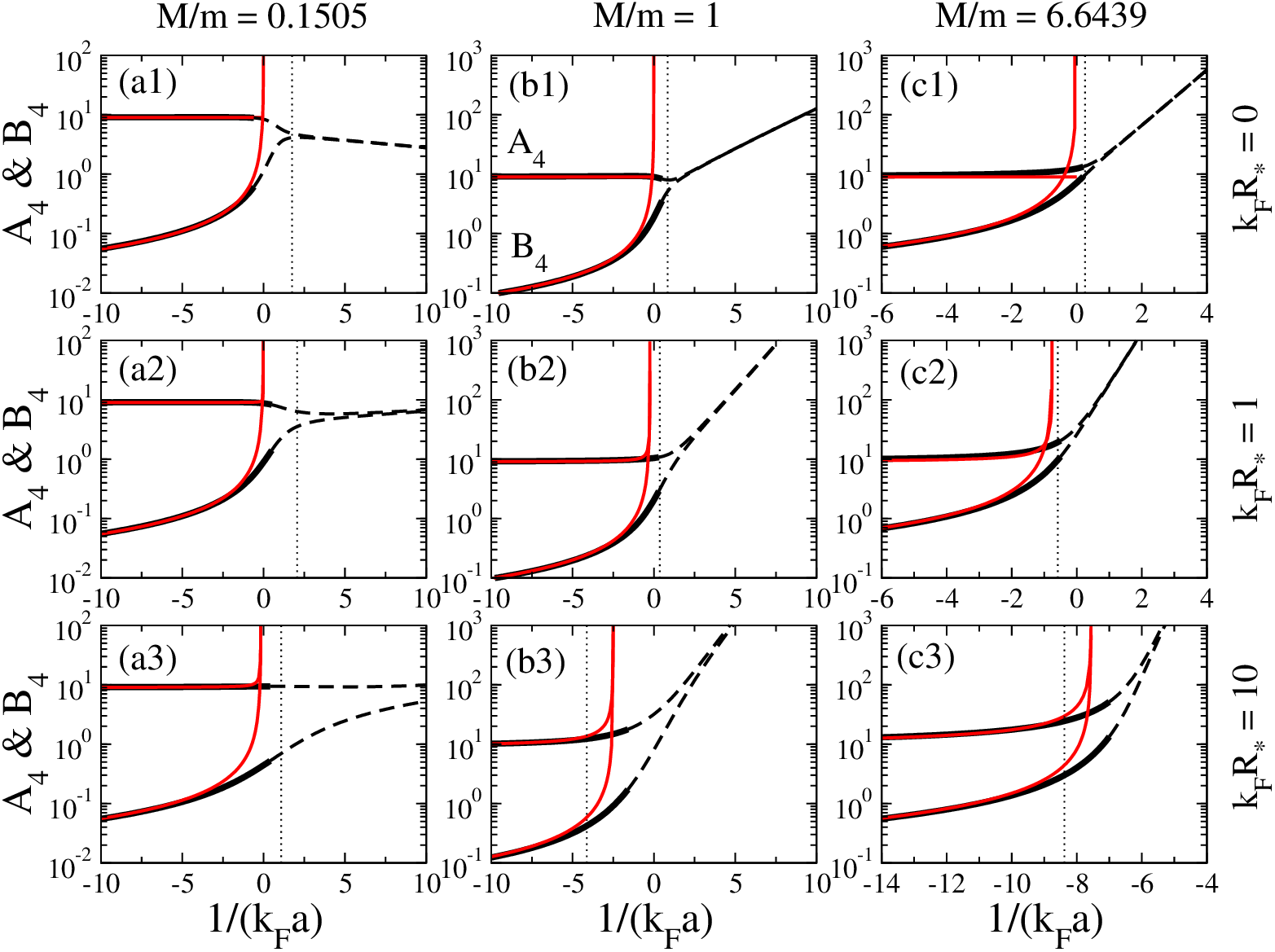}
\vspace{-0.5cm}
\caption{(Color online) Coefficients $A_4$ and $B_4$ in the asymptotic expansion (\ref{eq:g2large}) of the pair correlation
function, for various mass ratios $M/m$ (columns a,b,c)
and various Feshbach lengths (rows 1,2,3).
Upper and lower black lines (solid for $Z>1/2$, dashed for $Z<1/2$): Numerical evaluation 
of $A_4$ and $B_4$, respectively.
Red lines: Leading order analytical results (\ref{eq:A_4_a0}) and (\ref{eq:B_4_a1}), that 
diverge for $R_*>0$ when $s$ of Eq.~(\ref{eq:defs}) 
tends to unity, and for $R_*=0$ when $1/(\kf a)\to 0$. Vertical dotted lines: Polaron-to-dimeron crossing point.
}
\label{fig:b4}
\end{center}
\end{figure}

Eq.~(\ref{eq:g2large}) results from 
an asymptotic expansion of (\ref{eq:fq_def},\ref{eq:fk_def},\ref{eq:f_def}) 
in powers of $1/x$, obtained by repeated
integration by parts as in \cite{Holzmann}, always integrating the exponential function $e^{i\kk\cdot\xx}$
or $e^{\pm i\qq\cdot\xx}$ to pull out a $1/x$ factor
\footnote{One also uses the fact that, uniformly in the integration domain:
$\forall n_1,n_2,n_3,n_4 \in \mathbb{N}$, there exists $C_{n_1,n_2,n_3,n_4} \in \mathbb{R}^+$
such that $|\partial_k^{n_1} \partial_{u}^{n_2} \partial_q^{n_3} \partial_{u'}^{n_4} F(k,u;q,u') | 
\le C_{n_1,n_2,n_3,n_4}q^{n_2+n_4}$.}: 
\begin{eqnarray}
\label{eq:fq0def}
\!\!\!\!\! f_\qq(\xx)  &\!\!\!\!\!\sim\!\!\!\!\!& 
\frac{\mu/(2\pi^2\hbar^2)}{x^2 k_{\rm F} \mathcal{D}_\qq} \sum_{u=\pm 1} F(k_{\rm F},u;q,u')e^{ik_{\rm F}xu} \\
\label{eq:fk0def}
\!\!\!\!\! \tilde{f}_\kk(\xx) &\!\!\!\!\!\sim\!\!\!\!\!& \frac{-\mu \kf/(2\pi^2\hbar^2)}{x^2k^2 \mathcal{D}_{k_{\rm F}\ve_z}}\!\! 
\sum_{u'=\pm 1} \!\! F(k,u;k_{\rm F},u')e^{ik_{\rm F}xu'}, \\
\label{eq:I_large_x}
\!\!\!\!\! f(\xx) &\!\!\!\!\!\sim\!\!\!\!\!& \frac{-\mu/(8\pi^4\hbar^2)}{x^4\mathcal{D}_{k_{\rm F}\ve_z}}\!\!\!\! \sum_{u,u'=\pm 1} \!\!\!\!
F(\kf,u;\kf,u') e^{i\kf x (u-u')}.
\end{eqnarray}
Here $\ve_z$ is the unit vector along $z$, $u$ is the cosine of the angle between $\xx$ and $\kk$, 
$u'$ is the cosine of the angle between $\xx$ and $\qq$, and the function $F$ is defined as
follows:
\begin{multline}
\label{eq:grandF}
F(k,u;q,u')=\bigg[
-\frac{4m^2}{(m+M)^2}(1-u^2)(1-u'^2)\frac{q^2}{k^2} \\
+\left(
1+\frac{m-M}{m+M}\frac{q^2}{k^2}-\frac{2\mu}{M}\frac{q}{k}uu'-\Delta E_{\rm pol}\frac{2\mu}{\hbar^2k^2}
\right)^2
 \bigg]^{-1/2}.
\end{multline} 
Therefore, in Eq.~(\ref{eq:g2_bs}) the integrals containing $|f_\qq(\xx)|^2$ and $|\tilde{f}_\kk(\xx)|^2$ 
provide a $1/x^4$ contribution with an oscillating prefactor, as $f(\xx)$ does. 
In Fig.~\ref{fig:b4} we plot  $A_4$ and $B_4$ as functions of $1/\kf a$ for various values of the mass ratio $M/m$ 
and reduced Feshbach length $\kf R_*$. 

\section{In the weakly attractive limit}
Following Ref.~\cite{TrefzgerCastin}, we define for $a<0$
\be
\label{eq:defs}
s\equiv \frac{\mu}{m}\,(-aR_*)^{1/2}\kf,
\ee
and we take the limit $a\to0^-$ for fixed $s<1$ 
(which implies $R_*\to \infty$).
Then $\Delta E_{\rm pol}$ tends to $0^-$, so that the integral 
appearing in (\ref{eq:defD}) is bounded~\cite{TrefzgerCastin} and
\be
\label{eq:gD}
g\mathcal{D}_\qq\underset{a\to 0^-}{\to} 1-(sq/\kf)^2.
\ee
After integration over $\qq$ in Eq.~(\ref{eq:Epol}), we get
as in \cite{TrefzgerCastin}:
\be
\Delta E_{\rm pol} \underset{a\to 0^-}{\sim} \frac{\hbar^2 k_{\rm F}^2}{\mu} 
\frac{k_{\rm F} a}{\pi} \frac{1}{s^2}
\left[\frac{\arctanh s}{s} -1\right].
\label{eq:epol_lead}
\ee
Also, the interchannel coupling
amplitude $\Lambda$ scales as $(-\kf a)^{1/2}$, see Eq.~(\ref{eq:defct}),  and a systematic expansion of observables
may be performed in powers of $\kf a$, treating the interchannel coupling $\Lambda \hat{b}^\dagger \hat{u}
\hat{d}+\mbox{h.c.}$ within perturbation theory.
The terms neglected in the ansatz (\ref{eq:ansatz_polaron}) have an amplitude
$O(\Lambda^3)$, that is a probability $O(\Lambda^6)$. One can thus extract from
Eq.~(\ref{eq:defZ}) the exact value of $Z$ up to $\Lambda^4$:
\be
\label{eq:Zwal}
\frac{1}{Z} \underset{a\to 0^-}{=} 1+ \frac{m^2}{\mu^2}\left[c_1 \frac{\kf a}{\pi}
+c_2\left(\frac{\kf a}{\pi}\right)^2+O(\kf a)^3\right].
\ee
The first coefficient of the expansion (\ref{eq:Zwal}) is given by
\be
c_1 = -\left(\frac{1}{1-s^2}-\frac{\arctanh s}{s}\right),
\ee 
it originates from the second term in Eq.~(\ref{eq:defZ}), a closed-channel
contribution. This is why $c_1=0$ on a broad Feshbach resonance where $s=0$.
The second coefficient is
\begin{multline} 
\label{eq:c2}
c_2= \frac{1}{2} \frac{(1+\alpha)^2}{1-s^2} \frac{(s\arctanh s-\alpha\arctanh \alpha)}{s^2-\alpha^2} \\
-\frac{1}{2}\left[1+\frac{2m^2}{\mu^2}\left(\frac{\arctanh s}{s}-1\right)\right]
\left[\frac{1+s^2}{(1-s^2)^2} - \frac{\arctanh s}{s}\right] \\
+\frac{1}{2(1-s^2)} +\frac{s^2(1-\alpha)(1+\alpha)^2\arctanh\alpha}{(1-s^2)^2(s^2-\alpha^2)}\\
-\frac{[s^4+(1+\alpha)^2s^2-\alpha^2]\arctanh s}{2s(1-s^2)(s^2-\alpha^2)},
\end{multline}
where the mass contrast $\alpha=(M-m)/(M+m)$ also obeys $2\arctanh\alpha=\ln(M/m)$
\footnote{Contrarily to a first impression, $c_2$ has a finite limit when $s\to \alpha$.}.
The first term in Eq.~(\ref{eq:c2}) originates from the last term in (\ref{eq:defZ}),
an open channel contribution.  For a broad Feshbach resonance ($s=0$), it is non-zero,
whereas the sum of the other terms of (\ref{eq:c2}), originating from the closed channel, 
vanishes, and $c_2=\frac{\ln(M/m)}{1-m^2/M^2}$.

It is instructive to analyze the perturbative expansion in the exactly solvable limit
of $M/m\to +\infty$: The impurity can then be considered as a pointlike scatterer of fixed position,
for convenience at the center of a spherical cavity of arbitrarily large radius $R$, 
imposing contact conditions of scattering length $a$ and effective range $-2R_*$ on the fermionic
wavefunction \cite{Petrov}.  In the thermodynamic limit, 
one can then construct the Fermi sea of exactly calculable 
single-particle eigenstates in this scatterer-plus-cavity problem. As shown in \cite{Combescot} on a broad Feshbach resonance, 
the truncated ansatz (\ref{eq:ansatz_polaron}) provides a good estimate of $\Delta E_{\rm pol}$ 
for $M/m\to \infty$.
On the contrary, we find that it is qualitatively wrong for the quasiparticle residue: 
From Eq.~(\ref{eq:defZ}), it predicts a non-zero value of $Z$, whereas the exact $Z$ vanishes
for $M/m\to\infty$, which proves the disappearance of the polaronic character.
For an infinite mass impurity, indeed,  $Z$ is the modulus squared of the overlap 
between the ground state of the free Fermi gas and the ground state of the Fermi gas interacting with the
scatterer. This overlap was studied in \cite{Anderson}, and vanishes in the
thermodynamic limit, a phenomenon called the Anderson orthogonality catastrophe.
Satisfactorily, the perturbative expansion (\ref{eq:Zwal}) is able to detect this catastrophe:
\be
c_2 \underset{M/m\to\infty}{\sim} \frac{\ln(M/m)}{(1-s^2)^2}.
\ee
Such a logarithmic divergence with the mass ratio was already encountered in the context of
the sudden coupling of a Fermi sea to a finite mass impurity,
see the unnumbered equation below (4.3) in \cite{Toulouse}.
It originates from the first term of
(\ref{eq:c2}), that is from the last term of (\ref{eq:defZ}), where it is apparent
that $\int'\!d^3 k d^3q/(k^2-q^2)^2$ diverges logarithmically at the Fermi surface.

Turning back to a finite mass ratio $M/m$,
we now calculate the coefficients $A_4$ and $B_4$ in Eq.~(\ref{eq:g2large}) to leading order in $\kf a$.
The integrals appearing in Eq.~(\ref{eq:g2_bs}) contribute to these coefficients to second order in $\kf a$, see
Eqs.~(\ref{eq:fq0def},\ref{eq:fk0def}), since $1/\mathcal{D}_\qq$ scales as $g$ \footnote{The integrals over $\kk$ of the modulus squared
of (\ref{eq:fq0def}), and over $\qq$ for the modulus squared of (\ref{eq:fk0def}), have contributions scaling as $(\kf a)^2\ln (\kf a)$
that cancel in their difference.}. Also $Z/\pi_{\rm open}$ deviates from unity only to second order in $\kf a$,
see Eq.~(\ref{eq:def_rho_d}).
The leading order contribution to $A_4$ and $B_4$ is thus given by the interference term $f(\xx)$, and from Eq.~(\ref{eq:I_large_x}) 
we obtain
\bea
\label{eq:A_4_a0}
A_4 &\underset{a\to 0^-}{\to}& A_4^{(0)}=\frac{3 s^2}{1- s^2} \left(\frac{\arctanh s}{s}-1\right)^{-1}, 
\\
\label{eq:B_4_a1}
B_4 &\underset{a\to 0^-}{\sim}& -\frac{3}{2\pi}\left(1+ \frac{M}{m}\right) \frac{\kf a}{1-s^2}.
\eea
These analytical predictions are satisfactorily compared to the numerical evaluation
of $A_4$ and $B_4$  in Fig.~\ref{fig:b4}.

Taking again $M/m\to +\infty$, we find a severe divergence in the small-$\kf a$ expansion: The coefficient
of the term linear in $\kf a$ in $B_4$ diverges linearly with the mass ratio.
Instead, the coefficient of the term linear in $\kf a$ in $A_4$ (not given) 
diverges only logarithmically with the mass ratio 
$\sim -A_4^{(0)}\frac{\ln(M/m)}{\pi (1-s^2)}$. This suggests that, for an impurity of infinite mass, 
the asymptotic behaviour of $G(\xx)-1$ is no longer $O(1/x^4)$ and that it decreases more slowly.
To confirm this expectation, restricting for simplicity to a broad Feshbach resonance ($R_*=0$),
we have calculated the exact mean fermionic density in presence of a fixed pointlike scatterer, obtaining an expression equivalent
to the one of \S 2.2.2
of \cite{Giraud} and leading at large distances to 
\be
G(\xx) -1 \underset{x\to +\infty}{\sim} \frac{3}{2 (\kf x)^3}\, \mbox{Re}\, \left(e^{2i\kf x}\frac{\kf a}{1+i\kf a}\right).
\ee
As this $1/x^3$ law has a zero-mean oscillating prefactor, $G(\xx) -1$ has a non-diverging integral
over the whole space. {\sl Complement in augmented version:} the mean fermionic density in the presence of the fixed scatterer of scattering length $a$ placed at the origin of coordinates is $\rho+\delta\rho(\rr)$; for $a<0$ we obtain the expression
\begin{equation*}
\delta\rho(\rr) =\!\! \int_0^{k_{\rm F}}\!\! \frac{dk}{2\pi^2r^2} \left[\frac{k^2a^2}{1+k^2a^2} \left(\cos kr
\!-\!\frac{\sin kr}{ka}\right)^2\!\! -\sin^2 kr\right]
\end{equation*}
This gives $G(\xx)-1 = \delta\rho(\xx)/\rho$. Interestingly the sum rule (\ref{eq:regsom}) is no longer obeyed.
As the integral in (\ref{eq:regsom})  is no longer absolutely convergent, some care must be taken.
We find that $\delta\rho(\rr)$ contains a mean number of fermions within the ball of radius $R$
\begin{eqnarray}
N(R) &\equiv& \int_{r<R} d^3r\, \delta\rho(\rr) \nonumber \\
&=& \frac{a}{\pi} \int_0^{k_{\rm F}} dk\left[\frac{ka\sin(2kR)}{1+k^2a^2}
+\frac{\cos(2kR)-1}{1+k^2a^2}\right] \nonumber
\end{eqnarray}
that has, when $R$ tends to infinity, a nonzero limit determined by the nonoscillating bit in the integrand:
\begin{equation*}
\lim_{R\to +\infty} N(R) \underset{a<0}{=} \frac{\atan(-k_{\rm F}a)}{\pi}
\end{equation*}
{\sl End of complement.}

\section{Multiscale structure of $G(\xx)$}

In the weakly attractive limit, it is apparent from Eqs.~(\ref{eq:fq_def},\ref{eq:fk_def},\ref{eq:f_def})
that the functions $f_\qq$, $\tilde{f}_\kk$ and $f$ are of order one in $\kf a$, so that the leading order contribution
to $G(\xx)-1$ in Eq.~(\ref{eq:g2_bs}) originates from the interference term $\propto f(\xx)$ and is also of
order one:
\be
G(\xx) -1 \underset{a\to 0^-}{=} O(\kf a).
\ee
One then expects that, when $\kf x \gg 1$, $G(\xx)-1$ drops according to the  asymptotic $1/x^4$ law with 
a coefficient of order one in $\kf a$. This simple view is however infirmed by Eq.~(\ref{eq:A_4_a0}),
where $A_4$ has a non-zero limit $A_4^{(0)}$ for $a\to 0^-$. This suggests that the $1/x^4$ law
is only obtained at distances that diverge in the weakly attractive limit.

To confirm this expectation, one calculates $f(\xx)$ to first order in $\kf a$ for
a fixed $\xx$ [by using (\ref{eq:gD}) and neglecting
$\Delta E_{\rm pol}$ in the denominator of (\ref{eq:f_def})], then one takes the limit $\kf x \gg 1$.
As shown in the appendix, this gives
\begin{multline}
\!\!\!\lim_{a\to 0^-} \frac{G(\xx)-1}{\epsilon}\! \underset{x\to\infty}{=} \! \frac{A_4^{(0)}}{(\kf x)^4} 
\Big\{-\frac{M}{4 m} \cos(2\kf x) +\frac{s^2}{2(1-s^2)}  \\
+\frac{m}{2M} \Big[\gamma -\frac{3}{2} +\ln (2 \kf x M/m)\Big]\Big\}
+O\Big[\frac{1}{(\kf x)^5}\Big]
\label{eq:limsim}
\end{multline}
with the positive quantity $\epsilon \equiv -\Delta E_{\rm pol}/E_{\rm F} \ll 1$
and $\gamma\simeq 0.577\, 215$ is Euler's constant. 
For that order of taking limits, 
the oscillating bit still obeys a $1/x^4$ law, with the same coefficient $B_4$ as in (\ref{eq:B_4_a1});
on the contrary, the non-oscillating bit obeys a {\sl different} $\ln x/x^4$ asymptotic law 
(dotted line in Fig.~\ref{fig:g2}b), which shows
that the validity range of the $1/x^4$ law is pushed to infinity when $a\to 0^-$.

Remarkably, by keeping $\Delta E_{\rm pol}$ in the denominator of (\ref{eq:f_def}), one can obtain,
see the appendix,
an analytical expression for $G(\xx)-1$ that contains both the $\ln x/x^4$ 
and the $1/x^4$ laws as limiting cases, and that describes the crossover region with cosine- and sine-integral 
functions:
\begin{multline}
G(\mathbf{x})-1\! \underset{\kf x > 1}{=}\! \frac{A_4^{(0)} \epsilon}{(\kf x)^4}
\Big\{
\frac{m}{M} \Big[\Ci(\kf x \epsilon^{1/2}) -\frac{1}{2} \Ci(\kf x \epsilon)\Big]  \\
-\frac{1}{4} \kf^2 x^2 \epsilon \Big[
\Big(\frac{\pi}{2} -\Si(\kf x \epsilon/2)\Big)\sin (\kf x \epsilon/2)  \\
-\Ci(\kf x \epsilon/2) \cos (\kf x \epsilon/2) \Big] +O(1)\Big\}
\label{eq:multiscale}
\end{multline}
where the remainder $O(1)$ is a {\sl uniformly} bounded function of $\kf x>1$ and $\epsilon \ll 1$.
This formula satisfactorily reproduces the numerical results, see Fig.~\ref{fig:g2}b, where
$\epsilon\simeq 0.160$. It reveals that the pair correlation function
has a multiscale structure for a weakly attractive interaction, with three spatial ranges 
\footnote{The first relative deviation of $(\kf x)^4 [G(\xx)-1]$ from its $x\to\infty$ limit
is $-24/(\kf x \epsilon)^2 + \frac{m}{2M x} \sin (\kf x\epsilon)$. For $m\epsilon/M < 6$, 
this is $<10\%$ for $\kf x \epsilon > 16$. From a similar first-deviation analysis,
being in the logarithmic range actually requires $\kf x\epsilon^{1/2}< (\mu/M)^{1/2}$.}:
(i) the logarithmic range, $1 < \kf x < \epsilon^{-1/2}$, (ii) the crossover range, $\epsilon^{-1/2} < \kf x
< 16\epsilon^{-1}$, and (iii) the asymptotic range, $16\epsilon^{-1} < \kf x$. 
The logarithmic range is immediately recovered from $\Ci(u)=\ln u +O(1)$.
The $\epsilon^{-1/2}$ scaling of its upper limit is intuitively recovered if one assumes
that the relevant wave vectors in (\ref{eq:f_def}) obey $|\kk-\qq|\approx 1/x$: Neglecting 
$\Delta E_{\rm pol}$ with respect to $E_{\kk-\qq}$ in the denominator of (\ref{eq:f_def})
then indeed requires $\kf x \lesssim (M\epsilon/m)^{-1/2}$.

\section{Conclusion}
The Fermi polaron, composed of an impurity particle dressed by the particle-hole excitations
of a Fermi sea close to a broad or narrow Feshbach resonance with zero-range interaction, 
is a spatially extended object: 
The density perturbation induced by the impurity in the Fermi gas 
asymptotically decays as the inverse quartic distance,
with a spatially modulated component reminiscent of the Friedel oscillations. 
In the weakly attractive limit, $\kf a\to 0^-$ with $|a| R_*$ fixed,
where systematic analytical results are obtained, 
this density perturbation reaches its asymptotic regime over distances diverging as $1/(\kf^2 |a|)$
and exhibits at intermediate distances a rich multiscale structure.
This may have important consequences on the interaction between polarons \cite{interpol}: A polaron should indeed
be sensitive to the deformation of the underlying fermionic density profile induced by another polaron, since
the impurity forming the polaron is coupled to that fermionic density. The long-range nature of $G(\xx)-1$
suggests that the resulting interaction may be also long range.

\acknowledgments
We acknowledge financial support from the ERC Project FERLODIM N.228177. C. Trefzger acknowledges
support from a Marie Curie Intra European grant INTERPOL N.298449 within the 
7th European Community Framework Programme.

\section{Appendix}
{The integral over $\kk$ and $\qq$ in $f(\xx)$ reduces to a triple integral over $k$, $q$ and the angle $\theta$ between $\kk$ and $\qq$. Taking $\lambda=|\kk-\qq|$ rather than $\theta$ as
the variable, and changing the integration order, 
we get $f(\xx)=(-\mathcal{F}/x) \int_0^{+\infty}d\lambda \sin(\lambda x) \varphi(\lambda)$ with $\mathcal{F}=(1-s^2)A_4^{(0)}\rho\epsilon/2$,
\begin{equation*}
\varphi(\lambda)=\!\! \int_{\max(1-\lambda,0)}^{1} dq \int_{\max(\lambda-q,1)}^{\lambda+q} dk \frac{kq/(1-s^2 q^2)}
{\frac{m}{M}\lambda^2 + k^2 -q^2 +\epsilon}
\end{equation*}
and $1/\kf$ is the unit of length.
$\varphi(\lambda)$ is a $C^2$ function over $[0,2]$
and $[2,+\infty[$, with $\varphi(0)=0$, $\varphi''(0)=1/[\epsilon (1-s^2)]$,
but with a jump $J=\varphi''(2^+)-\varphi''(2^-)\underset{\epsilon\to 0}{=} M/[4m(1-s^2)]$.
Triple integration by parts over each interval gives
\begin{equation*}
f(\xx) = \frac{\mathcal{F}}{x^4} \Big[J \cos 2x +\varphi''(0) +\int_0^{+\infty} d\lambda\,
\varphi^{(3)}(\lambda) \cos \lambda x\Big].
\end{equation*}
The contribution of $J$ reproduces Eq.~(\ref{eq:B_4_a1}).
We find that $\varphi(\lambda)$ varies at three scales, $\epsilon$, $\epsilon^{1/2}$ and $\epsilon^0$.
For $0<\lambda<\epsilon^{3/4}$, we use the scaling $\lambda=\epsilon t$ and expand $\varphi^{(3)}(\lambda)$ 
in powers of $\epsilon$ at fixed
$t$. For $\epsilon^{3/4}<\lambda<\epsilon^{1/4}$, we use the scaling $\lambda=\epsilon^{1/2} u$ and expand 
$\varphi^{(3)}(\lambda)$ at fixed $u$. For $\epsilon^{1/4} < \lambda <1$, we directly expand at fixed $\lambda$. With $\eta=m/M$, this gives
\begin{eqnarray*}
\varphi^{(3)}(\lambda)\!\!\! &=&\!\!\! -\frac{4/(1-s^2)}{\epsilon^2 (1+2t)^3} - \frac{\eta Y(t-1)/t+O(2t+1)^{-3}}
{2\epsilon (1-s^2)} +\ldots \\
\varphi^{(3)}(\lambda)\!\!\! &=&\!\!\! - \frac{(1-\eta u^2)(1+4\eta u^2 + \eta^2 u^4)}
{2\epsilon^{1/2}u^3 (1+\eta u^2)^2 (1-s^2)} + \frac{3+O(u^4)}{4u^4 (1-s^2)}+\ldots \\
\varphi^{(3)}(\lambda)\!\!\! &=&\!\!\! \Big[\frac{\eta}{2\lambda (1-s^2)} +O(\lambda^0) \Big]
+\frac{\epsilon [1+O(\lambda)]}{2\lambda^3 (1-s^2)}+\ldots
\end{eqnarray*}
where $Y$ is the Heaviside function, and the $O(\,)$ apply for $t\in \mathbb{R}^+$, $u\in \mathbb{R}^+$
and $\lambda\in [0,1]$, respectively. If one needs the integral of 
$\varphi^{(3)}(\lambda) \cos(\lambda x)$ 
up to an error $O(1)$ {\sl uniformly} bounded
in $x$ and $\epsilon$, one can apply several simplifications over each interval. E.g., for $\epsilon^{1/4}
< u <1$, one can approximate $\varphi^{(3)}(\lambda)$ by an expansion in powers of $u$, and for $1<u<\epsilon^{-1/4}$,
by an expansion in powers of $1/u$. Adding contributions of all intervals, we concatenate them
by pairs, and further noting that $J=O(1)$, $\int_{\epsilon^{-1/4}}^{\epsilon^{-1/2}} \frac{dt }{\epsilon}
|t^{-3}-\frac{3}{2} t^{-4} -(t+1/2)^{-3}|=O(1)$, $\int_{\epsilon^{-1/2}}^{+\infty} \frac{dt}{\epsilon} 
t^{-3} = O(1)$, $\int_1^{+\infty} d\lambda |\varphi^{(3)}(\lambda)|=O(1)$, we get
\begin{multline*}
G(\xx)-1 = \frac{A_4^{(0)}\epsilon}{x^4}\Big\{-\frac{1}{\epsilon}+ \int_0^{+\infty} \frac{dt}{2\epsilon}
\frac{\cos(x \epsilon t)}{(t+1/2)^3} \\
+ \eta\Big[\int_\epsilon^{\epsilon^{1/2}} \frac{d\lambda}{2\lambda} \cos(\lambda x)
- \int_{\epsilon^{1/2}}^{1}  \frac{d\lambda}{2\lambda} \cos(\lambda x)\Big] + O(1) \Big\}.
\end{multline*}
Explicitly evaluating the integrals gives Eq.~(\ref{eq:multiscale})
\footnote{We have checked that the leading correction to (\ref{eq:gD}) gives, as expected, a contribution
also $O(1)$ to the expression in between curly brackets in (\ref{eq:multiscale}).}.

Finally, to obtain (\ref{eq:limsim}), one omits $\epsilon$ in the denominator of $\varphi(\lambda)$,
so that it is no longer $C^2$ at the origin: 
$\varphi''(\lambda)=\{\eta[\ln(\eta\lambda/2) +3/2] -s^2/(1-s^2)\}/[2(1-s^2)]+O(\lambda)$.
We thus locally split $\varphi''(\lambda)$ as the sum of a singular part $\propto \ln\lambda$ and a $C^\infty$ 
function.
The only trick is then to take, in the last triple integration by parts over $\lambda$ and in the bit
involving the singular part, $(1-\cos\lambda x)/x$ 
as a primitive of $\sin \lambda x$.
}

\end{document}